# In-house beam-splitting pulse compressor with compensated spatiotemporal coupling for high-energy petawatt lasers

**Jun Liu**[1,2,3,4], **Xiong Shen**[1], **Zhe Si**[1,2], **Cheng Wang**[1], **Chenqiang Zhao**[1], **Xiaoyan Liang**[1,2,3], **Yuxin Leng**[1,2,3], **Ruxin Li**[1,2,3,5]

[1]*State Key Laboratory of High Field Laser Physics, Shanghai Institute of Optics and Fine Mechanics, Chinese Academy of Sciences, Shanghai 201800, China*
[2]*University Center of Materials Science and Optoelectronics Engineering, University of Chinese Academy of Sciences, Beijing 100049, China*
[3]*CAS Center for Excellence in Ultra-intense Laser Science (CEULS), Shanghai Institute of Optics and Fine Mechanics, Chinese Academy of Sciences, Shanghai 201800, China*
[4] *jliu@siom.ac.cn*
[5] *ruxinli@mail.siom.ac.cn*

**Abstract:** One of the most serious bottleneck on achieving kilojoule-level high-energy petawatt (PW) to hundreds-petawatt (100PW) lasers with ps to fs pulse duration is the requirement of as large as meter-sized gratings in the compressor so as to avoid the laser-induced damage to the gratings. However, this kind of meter-sized grating with high quality is hard to manufacture so far. Here, we propose a new in-house beam-splitting compressor based on the property that the damage threshold of gratings depend on the pulse duration. The new scheme will simultaneously improve the stability, save expensive gratings, and simplify the size of compressor because the split beams share the first two parallel gratings. Furthermore, based on the fact that the transmitted wavefront of a glass plate can be much better and more precisely controlled than that of the diffraction wavefront of a large grating, then glass plates with designed transmitted wavefront are proposed to compensate the wavefront distortion introduced by the second, the third gratings, and other optics in-house such as the beam splitter. This simple and economical method can compensate the space-time distortion in the compressor and then improve the focal intensity, which otherwise cannot be compensated by the deformable mirror outside the compressor due to angular chirp. Together with multi-beams tiled-aperture combining scheme, the novel compressor provides a new scheme to achieve high-energy PW-100PW lasers or even exawatt lasers with relatively small gratings in the future.

## 1. Introduction

Kilojoule-level high-energy laser facilities with peak power from petawatt (PW) to hundreds-PW are opening up lots of important research fields in ultrahigh intensity sciences, such as laboratory astrophysics, particle acceleration, nuclear physics, and nonlinear QED etc. [1-2]. Over the past decade, tens of lasers in PW-level with pulse duration from tens femtosecond to hundreds femtosecond had been demonstrated all over the world for these exciting research fields [3]. So far, PW laser with the highest pulse energy was achieved by LFEX from Osaka University, in which the compressed output pulse energy is about 2000 J with a peak power of 2PW [4]; while 339 J high energy was also obtained in a Ti:sapphire chirped-pulse amplifier by SIOM in 2018 [5]. Several 10 PW laser facilities all over the world are in construction: the SULF facility from SIOM in China (SULF-10 PW), the ELI-Nuclear Physics (ELI-NP) in Romania, the APPOLON from CNRS in France, the upgrading Vulcan laser located at the Central Laser Facility (CLF) in United Kingdom, and PEARL-10 PW at the Institute of Applied Physics of the Russian Academy of Sciences in Nizhny Novgorod (Russia) [6-8]. Moreover, even 100 PW-level or sub-exawatt laser facilities are also being commissioned, such as SEL

(Station for Extreme Light, China), OPAL (Optical Parametric Amplifier Line, USA) and XCELS (Exawatt Center for Extreme Light Studies, Russia), and ELI, among which the SEL facility had been started up in 2018 [9-12].

In all these laser facilities, chirped-pulse amplification (CPA) or optical parameter chirped-pulse amplification (OPCPA) are the key techniques to achieve high-energy output which had been significantly improved during past years. To obtain PW to 100-PW high-energy laser pulse, the amplified high-energy chirped pulse has to be compressed back to ultrashort pulse duration typically by using pulse compressor based on parallel diffraction gratings. It needs to be noted that the laser-induced damage in gratings is tightly related to the pulse energy and beam size on the surface of gratings. Pulse duration and wavelength will also affect the damage threshold depend on the type of gratings [13]. To satisfied kilojoule-level femtosecond pulses output, that means muti-PW with hundreds femtosecond or hundreds-PW with tens femtosecond, meter-sized gratings are urgently required to avoid the laser-induced damage on the surface gratings, especially the final grating of the whole compressor system due to the shortest pulse duration. However, gratings with large size and other needed nice specifications simultaneously are difficult or infeasible to be fabricated right now, and hence their cost would be very high. Moreover, a thicker substrate with smaller thermal expansion coefficients is necessary to maintain the necessary wavefront quality for a larger aperture grating which leading to a heavy weight and thereby difficult to handle during maintenance.

To solve this problem, tiled grating (or mosaic grating) method is proposed in 2004 for PW lasers [14-15] based on the arrayed grating scheme demonstrated in 1998[16]. It is because small gratings are easier to handle and keep their wavefront quality. So far, several laser facilities achieved PW output by using this grating phasing method. Tiled-grating method has been used to the second and third gratings to extend the spectral bandwidth or beam size of input lasers. Furthermore, even all the four gratings in the Fig. 1(a) will be replaced by tiled-gratings to extend its size and increase the input pulse energy. Two meter-sized gratings or three gratings with half-meter size are tiled together to replace all four gratings of a typical four-grating compressor in Omega-EP (Laboratory for Laser Energetics (LLE), Rochester) [17]. However, when consideration of the groove of the grating, each grating owns six degrees of freedom for tiled-grating methods: piston and shift, groove spacing and tilt, rotation and tip. Then, the grating is very sensitive due to six degrees of freedom that needs to be precisely controlled. Recently, an object-image-grating self-tiling compressor is proposed to reduce these sensitivities [18].

Besides tiled grating, tiled-aperture beam combining method was also proposed in 2006 [19]. The advantage of this scheme is that less parameters need to be controlled in comparison to that of the grating phasing [20]. This tiled-aperture combining method were extensively studied theoretically and experimentally by using tens femtosecond with small beam size recently [21-24]. In 2017, a 1.15 PW PETAL with 700 fs and 850 J energy was obtained in a tiled-aperture beam combining compressor [25]. Hundreds-PW laser facilities proposed by ELI and XCELS are both based on tens of separated amplifiers and compressors together with the tiled-aperture combing scheme. The relative long optical paths will increase the disturbance influences from vibration, airflow or temperature. Moreover, too many independent optical tables and optical holds will also bring large pointing and piston phase instability among different amplified and compressed beams which will make the final precise beam phasing very difficult.

In this paper, we propose a new compressor scheme for high-energy PW lasers when the pulse energy can be amplified high-enough to kilojoule-level with single beam. In the designed compressor, the amplified laser beam is split into two or more equal parts on pulse energy in between the second and the third gratings of a typical four-grating compressor. Every split parts are then compressed by two parallel grating pairs separately for further pulse compression which will avoid the damage of every final grating. To compensate the in-house wave-front distortion with angular chirp introduced by the second grating, the third grating, and the splitter in compressor which otherwise cannot be compensated by the deformable mirror outside the compressor, special designed glass plates with a conjugated transmitted wave-front which will well compensate the above wave-front distortion are proposed between the second and the third gratings of a typical four-grating compressor. This new optical design will simultaneously improve the stability, simplify the compressor, and save the expensive gratings because the split beams share the first two parallel gratings. At last, kilojoule-level PW to 100PW laser pulses can be obtained by tiled-aperture combining the separately compressed beams.

## 2. Four-grating compressor

For PW laser facilities, parallel diffraction grating pair is usually used as the final compressor stage for pulse compression which can introduce enough negative chirp with large beam size. The optical setup of a typical four-grating compressor is shown in Fig.1 (a). They are composed of two sets of parallel diffraction grating pairs which are arranged tending to be mirror images of each other. The beam size on the second and third gratings are expanded in the diffraction plan due to angular dispersion of the input beam. The beam size on the first and the last gratings are relatively small, then the laser-induced damage risk mainly occurred in these two gratings. As for the first grating, the damage risk comes from the highest pulse energy it should bear, while the damage risk for the last grating is due to high energy and the shortest pulse duration.

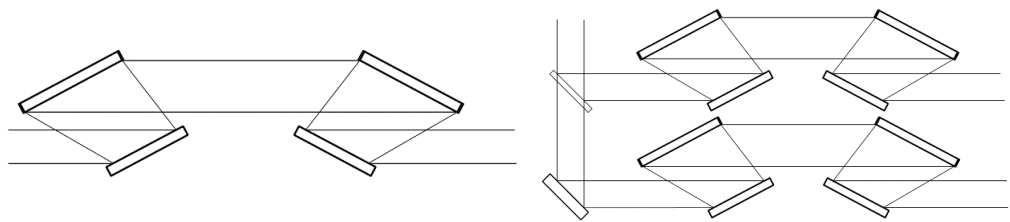

Fig.1 (a). The diagrammatic sketch of a typical four-grating compressor. (b) The diagrammatic sketch of an outside beam-splitting pulse compressor.

As for hundreds-PW lasers proposed by ELI and XCELS, tens of four-grating compressor are used after tens of separated amplifier [11-12]. For SEL (Station for Extreme Light, China) and OPAL (Optical Parametric Amplifier Line, USA) system, the pulse energy will be amplified to several thousands joules with single beam directly. Then, it needs gratings with both large size and high damage threshold simultaneously if single four-grating compressor is used. To the best of our knowledge, there is no usable grating satisfy these requirements up to now. It would be very natural that the beam should be split into several beams by using either wavefront beam split method [25] or amplitude beam split method shown in Fig1. (b), where every split beam is compressed separately. The diffraction effect would not be neglected for wavefront beam split method due to long distance after the beam splitting. And it need really a big vacuum chamber for separated compressors in both kinds of beam split methods. Then, how to simplify this beam split is an important issue should be solved. In the PETAL system with 700 fs and 850 J, two compressor are used, where the beam is split into several beams by in-house wavefront dividing method [25].

## 3. Scheme of in-house beam-splitting compressor

We noticed that many experimental results had shown that the laser-induced damage threshold of gratings decreasing with the shortening of pulse duration [13]. For gold-coated grating which own broadband bandwidth suit for tens femtoseconds pulse, the damage threshold of a chirped nanosecond pulse will be about three times higher than that of the compressed femtosecond pulses [13]. As for dielectric coated gratings, the damage threshold ratio between nanosecond chirped pulses and compressed femtosecond pulses will reach as high as few hundreds [13]. Consideration of the diffraction efficiency of the first three gratings (72.9% in all for 90% diffraction efficiency of every grating), the damage threshold energy on the first input grating can support a pulse energy that could be two times higher than the damage threshold of the last output grating even for a gold-coated grating. Then, we can introduce a beam splitter (BS) between the second grating (G2) and the third gratings (G3, G5) to split the beam into two beams and then guide them into two sets of parallel diffraction grating pair for the final pulse compression, as shown in Fig.2. In this way, two four-grating compressors share the first parallel diffraction grating pair, which will save two expensive gratings in comparison to the design by using beam splitting before compressor shown in Fig 1(b). Moreover, all the influences of the first gratings pair, including spectral phase, piston phase, and beam pointing stability, to the tiled-aperture combing are reduced which makes the final tiled-aperture combing process more easy and reliable. In another word, the stability of every compressor can be improved by 2 times because the first and the second pair of gratings are identical in a typical four-grating compressor. To make both parts have nearly the same spectral phase, the reflective part can pass through an added same thick glass plate so as to make a dispersion balance of both beams. Furthermore, the transmitted wavefront of both the splitter and the compensated glass plate (CP) can be custom designed and specially manufactured on the backside of these plates so as to compensate the wavefront distortion introduced by the second grating, the third grating and the optics in between in these typical four-grating compressors. Moreover, the rotation of the added glass plate can also be used to precisely tune the time delay of the two separated beams for tiled-aperture combing. Since a beam splitter and a glass plate are added in the novel compressor scheme, the influences of the beam splitter and the glass plate will be analysed and discussed in detail as following.

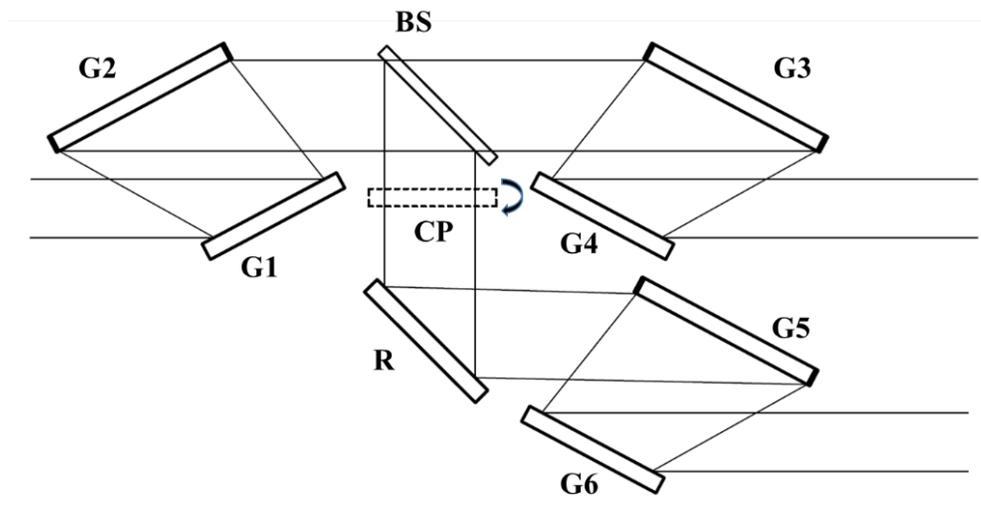

Fig.2. The sketch of proposed in-house beam-splitting pulse compressor. G1-G6: diffraction gratings, BS: beam splitter, CP: compensated plate, R: reflective mirror.

## 4. Spatial mode evolution in the compressor

The influence of the beam splitter between the second grating and the third grating in the compressor is tightly related to the spatial-temporal properties of the beam. Then, the spatial profile evolution before the beam splitter is studied through a simulation analysis. During the simulation, the input beam to the compressor is an tenth orders super-Gaussian beam with a size of 500mm×500mm. The chirped pulse duration is about 4 ns with the spectral range from 810 nm to 1010 nm and a Gaussian profile. The gratings own a line density of 1480 lines/mm, the incident angle is 62 degrees, and the central distance between the first and second grating is 1.3 m.

As for high-energy output laser, the spatial intensity usually has relatively large modulation mainly due to the intensity modulation of the pump laser. Then the laser beam before the compressor used in the simulation is shown in Fig. 3(a) which has been introduced a random intensity modulation of 2 times in the spatial domain. Owning to the angular dispersion of the first grating pair, the input beam is extended in the horizontal direction according to the wavelength. The second grating is used to collimate the angular dispersive beam introduced by the first grating into a plane beam. As a result, the spatial profile of the laser beam on the second grating will be well smoothed in comparison to that of the input beam, as shown in Figure 3(b). This is because the input laser owns a large beam size and a relative high temporal chirp to nanosecond, the intensity of every point on almost all part of the beam on the second grating is a sum or integral of the intensity of hundreds millimeter line of the input beam at different wavelengths. In the simulation, for an about 4 ns chirped laser pulse with spectral range from 810 nm to 1010 nm, an input beam spot in the first grating will be extended to an about almost 440 mm width line on the second grating. In the simulation, the beam size is enlarged to about 670 mm in the angular dispersive direction. Furthermore, the intensity modulation of the spatial profile after the second grating is well smoothed from original 2 times to about 1.1 times. At the same time, the pulse duration after the second grating is compressed to about half that of the input laser pulse, which is still in nanosecond level. As a result, the nonlinear effects induced by the splitter or the compensate plate which is located between the second and third gratings would be neglected owing to relative low intensity owing to this nanosecond chirped pulse, smoothed spatial profile and enlarged beam size.

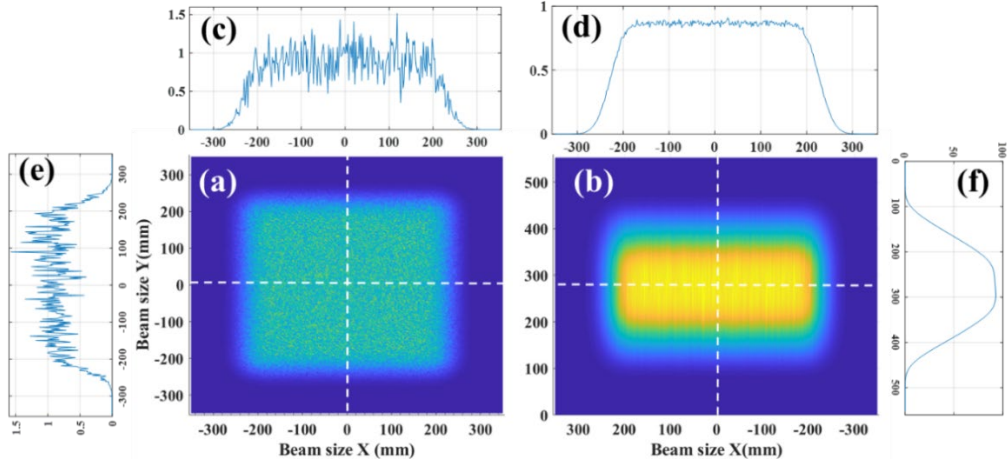

Fig.3. (a). The input beam before the compressor with about 2 times spatial intensity modulation randomly. (b) The simulated beam after the second grating or before the beam splitter. The simulation parameters are shown in the main text. (c)(e) and (d)(f) are the crossing beam profiles around the central of the input beam and the output beam(marked with white crossing dash line), respectively.

## 5. The influence of the added optics

In the novel design, the main influence is coming from the added optics in between the second and third gratings, they are beam splitter, compensated plate, and reflective mirror, then the influence of which will be analysed in both temporal domain and spatial domain. Actually, these beam splitter and reflective mirror are also needed in the outside beam-splitting pulse compressor shown in Fig. 1(b). The difference between these two designs is that the different located position in the compressor. From above analysis, the spatial modulation is smoothed and the beam size is enlarged, while there is angular chirp for the in-house beam-splitting pulse compressor in comparison to that of the outside beam-splitting one.

*5.1 Time domain*

In the time domain, the pulse duration for an ultrashort laser pulse is mainly decided by its spectral phase. As for the linear influence, the beam splitter will introduce the same fixed dispersion to the laser pulse for both in-house beam-splitting and outside beam-splitting pulse compressor. For the reflective beam of the in-house design, the compensated plate also own the same material and the same effective thickness to that of the beam splitter which located at 45 degrees to make the added dispersion on both reflective and transmitted sides equal. This additional spectral phase due to material dispersion of the beam splitter or the compensated plate can be well compensated by the four-grating compressor through slightly tuning the distance of the parallel grating pair. The reflective mirror R can be gold-coated with broad reflective spectral range and negligible dispersion. Moreover, the spectral phase differences between the two splitting beams are only decided by the differences of grating pairs of G3-G4 and G5-G6 for the in-house design. There are less parameters to tune to obtain equal spectral phase owing to the share of G1-G2 grating pair, then only the grating distance or angle of G3-G4 pair and G5-G6 pair need to be tuned. As a result, in comparison to that of outside beam-splitting design shown in Figure 1(b), the novel in-house beam splitting design would be easier to obtain the same spectral phase for the two splitting beams. Then, it will help to achieve an efficiency tiled-aperture combining which is affected by the spectral phase differences of independent beams [20, 26].

*5.2 Spatial domain*

In spatial domain, the wavefronts of both reflective and transmitted beams will be affected by the beam splitter. In the compressor, the wavefront distortion introduced by the first and fourth gratings can be compensated by using a deformable mirror outside the compressor. The wavefront distortion introduced by the optics in between the second and third gratings can be equal to an equivalent value of wavefront aberration added to the second or the third diffraction gratings. In the compressor, the wavefront aberration on the horizontal direction (angular dispersive direction) will introduce different spectral phase distortions to the compressed pulse on every spatial spots, while the wavefront aberration on the vertical direction may introduce wavefront on the spatial mode. Then, the same as the influences of the second or the third diffraction gratings, the wavefront aberration introduced by the additional optics in house will affect not only the pulse duration but also the final focal intensity and the temporal contrast [27-31]. It should be noted that this kind of wavefront distortion cannot be compensated by the deformable mirror outside the compressor.

In reality, it is hard to obtain a nice wavefront for a meter-sized grating. It had been shown that this wavefront distortion of gratings will decrease the focal intensity badly both in pulse duration and focal beam diameter [27-31]. Usually, the surface wavefront induced by the beam splitter, or reflective mirror are much better than that of the gratings. These in-house optics of the compressor with high quality of reflective and transmitted wavefront would help reduce their influences to the in-house wavefront aberration. Anyway, the additional wavefront distortion, especially the serious wavefront distortion induced by the second and the third gratings have to be compensated to achieve a nice focal intensity and nice temporal contrast, which will be discussed in the following section.

### 5.3 Nonlinear effects

Except for the linear effects such as wavefront aberration and material dispersion, it will introduce some nonlinear effects both in time domain and spatial domain when a high peak power laser pulse pass through a thick glass plate. The transmitted beam will pass through the beam splitter, while the reflective beam will pass through a compensated plate with almost same effective thickness. Then, nonlinear optical effects in the glass plate of both split beams will be analysised as following.

The B integral (nonlinear phase distortion) is usually used to characterize the nonlinear influences in ultrahigh intense laser system. In comparison to the scheme in Fig1.(b), in which the beam splitter is located before the compressor, the pulse duration before the in-house beam splitter is compressed by about 2 times while the beam size is also increased by above 2 times. When consideration of the energy loss in the first two diffraction gratings, the laser intensity on this in-house beam splitter is even smaller than that of the scheme in Fig1. (b). Furthermore, the pulse energy is split into two beams by the beam splitter. As a result, the pulse laser intensity is divided by two again for both the reflective and the transmitted beams. Here, the additional B integral induced by the beam splitter or the compensated plate can be expressed as $B = 2\pi/\lambda \int n_2 I(z) dz$, where $\lambda$ is the laser central wavelength, $n_2$ is the nonlinear refractive index of the glass plate, I is the laser intensity. From the spatial mode evolution analysis in section 4, we can see that the spatial mode is great smoothed in the compressor which will weak small-scale self-focusing heavily. Then, we can assume the B integral equal to $B = 2\pi n_2 IL/\lambda$ approximatively, where $L$ is the optic length of the glass plate. Take the example of a 100PW laser facility, where gold-coated gratings with broadband spectral range are used in the compressor, the energy intensity for the first input grating can be set to about 300 mJ/cm$^2$ to avoid the grating damage if assumption about 2 times of spatial intensity modulation of the input beam. The beam area on beam splitter in the horizontal direction would be almost the same that of on the first grating. As a result, the energy intensity on the beam splitter should be 240 mJ/cm$^2$ when considering the diffraction efficiency of the first pair of grating is about 80%. Assuming the chirped pulse duration of the input beam is 4ns, the pulse duration before the in-house beam splitter should be about 2ns. Then, the power intensity before the beam splitter should be 120MW/cm$^2$. For a half beam splitter with 45 degrees located, the power intensities of the reflective and the transmitted beams should be about 43MW/cm$^2$. In the case, the B integral is calculated according to the thickness of the beam splitter, as shown in **Figure 4,** where $n_2$ is about $2.7 \times 10^{-16}$cm$^2$/W at 910 nm for fused silica glass. Even for a 150 mm thickness glass beam splitter with a 45 degrees incident angle located, the B integral is calculated to be about 0.017 which is far smaller than 1. Then the nonlinear influence due to the B integral induced by the in-house beam splitter or the compensated plate is very small which can be neglected.

In comparison to that of the scheme in Fig1. (b), the spatial intensity distribution is smoothed in the horizontal direction in this novel in-house design, which means the high-order phase difference induced by the inhomogeneous intensity modulation would be smaller and can be ignored, the small-scale self-focusing effect in the beam splitter will also be restrained and is ignorable owning to the smoothed top-hat spatial profile and low intensity.

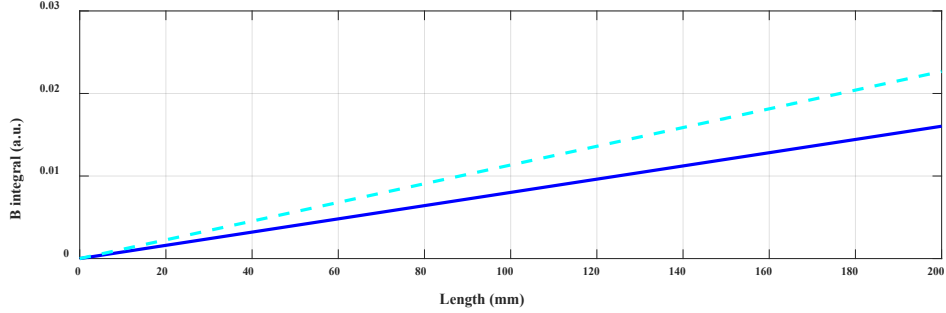

Fig.4. The calculated B integral according to the thickness of the beam splitter. Dash line and solid line mean 45 degrees and 0 degree incident angle to the incident beam, respectively.

## 6. Compensation of space-time coupling

According to many researches [27-32], ultrahigh intense laser with large beam size will suffer spatiotemporal coupling effects which will decrease the focal intensity badly. Even the added optics own relative nice wavefront, the diffraction wavefront distortion from the gratings with large size in the compressor would be large enough that need to be compensated [28, 31]. Although, precompensated compressor with small size had been proposed [33], it is not a direct method in the compressor and can only compensate spatiotemporal distortion induce by single compressor. It cannot work in this in-house beam-splitting compressor because there are two compressors, and then two in-house wavefront distortion need to be compensated independently. To compensate the spatiatemporal coupling effects induced by the second, the third gratings and the added optics between them, we propose a simple method to compensate the wavefront distortions by adding a compensated glass plates between the second grating and the third grating. This idea based on a well-known fact that the transmitted wavefront can be better controlled in comparison to that of reflective wavefront and diffraction wavefront for optics with large size. The optical setup of the principle idea is shown in Figure 5(a), where the back-side of the compensated plate CP is specially designed and manufactured, which is used to compensate the diffraction wavefront distortions induced by grating G2 and G3 in a typical four-grating compressor. The diffraction wavefront of G2 and G3 can be expressed as $D\_G2(x, y)$ and $D\_G3(x, y)$, while the transmitted wavefront of CP can be expressed as $T\_CP(x, y)$. Theoretically, the equation $T\_CP(x, y) = -(D\_G2(x, y) + D\_G3(x, y))$ is satisfied in perfect condition. It should be noted that this method can only compensate the static wavefront distortion due to optics deformation, and it cannot compensate the dynamic wavefront distortion due to heat-induced grating deformation because of high repetition rate or high average power [31]. Fortunately, almost all high-energy petawatt lasers operated at low repetition rate so far in which the wavefront distortion in the compressor can be well compensated by using a simple and special designed compensated plate.

As for the in-house beam-splitting pulse compressor shown in Fig. 2, the in-house wavefront distortion of the reflective split beam consist of the diffraction wavefront abbreviation of the second grating G2 and the fifth grating G5, the reflective wavefront of the beam splitter BS and reflective mirror R, and the transmitted wavefront of the compensated plate CP. Except for the usage of dispersive compensation of the beam splitter BS, the compensated plate CP is also used to compensate the in-house wavefront distortion of the reflective split beam. What we need to do are several steps: 1) measuring and achieve the diffraction wavefront aberration of gratings G2 and G5, which are $D\_G2(x, y)$ and $D\_G5(x, y)$; measuring and achieve the reflective wavefront aberration of BS and R, which are $R\_BS(x, y)$ and $R\_R(x, y)$; 2) design and write special wavefront phase to the surface of CP according to the achieved wavefront aberration value which equals to $D\_G2(x, y)+D\_G5(x, y)+R\_BS(x,$

y)+R_R(x, y), so as to the transmitted wavefront of the CP, which is T_CP(x, y), can compensate all the wavefront distortion of the reflective split beam. It means T_CP(x, y) =-(D_G2(x, y) +D_G5(x, y) +R_BS(x, y) +R_R(x, y)); 3) put the special designed CP to the compressor at the planned position.

As for the transmitted split beam shown in Fig. 2, the in-house wavefront distortion consist of the diffraction wavefront abbreviation of the second grating G2 and the third grating G3 which are D_G2(x, y) and D_G3(x, y), and the transmitted wavefront of the beam splitter BS, T_BS(x, y). Here the beam splitter can be used as the wavefront compensated plate simultaneously. After obtained the wavefront aberration value of G2 and G3, specially designed wavefront phase was written to the surface of BS on the back-side so that the transmitted wavefront of BS exactly compensate the diffraction wavefront aberration of G2 and G3. It means T_BS(x, y) =-(D_G2(x, y) +D_G3(x, y)).

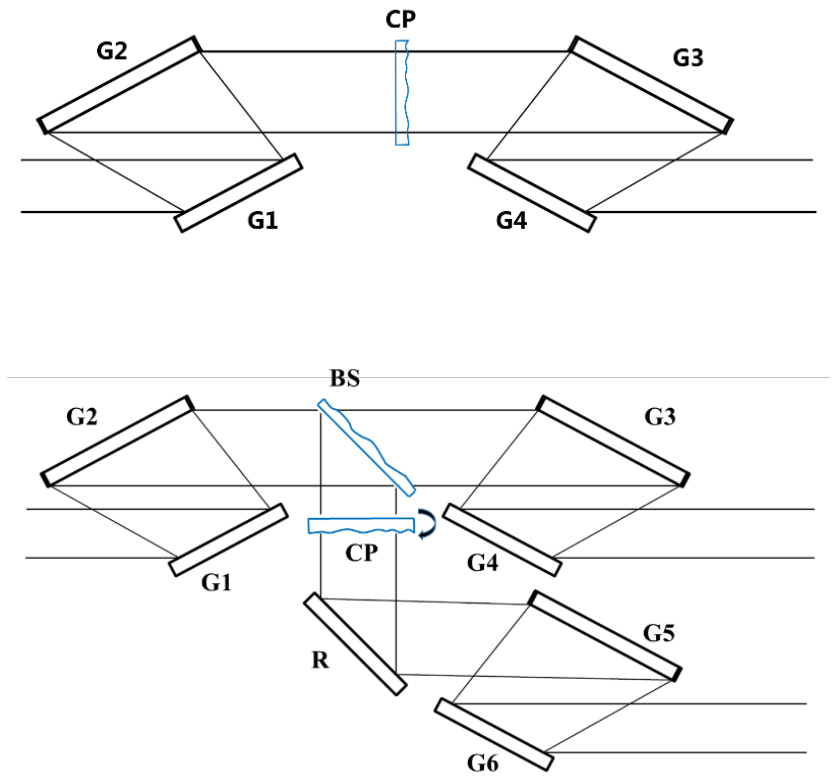

Fig.5 (a) The diagrammatic sketch of a typical four-grating compressor with a compensated plate CP; (b) The sketch of proposed in-house beam-splitting pulse compressor together with spatiotemporal coupling compensated plate CP.

With this simple method, the spatiotemporal effects of both the reflective beam and the transmitted beam are well compensated. Actually, one beam splitter without compensated plate can also worked in principle, where special designed phase on the front surface is used to compensated the wavefront distortion of the reflective beam, and special designed phase on the back surface for the transmitted beam. However, with another compensated plate CP, the manufacture of the BS is easier, and the dispersion of the BS can be balanced, the rotation of it can also be used to tune the time-delay of the reflective and the transmitted beam precisely.

According to the analysis previously, the laser spectra for a 100 PW laser can be extended from one spot to more than 400 mm on the second grating. Some experiments had shown that the diffraction wavefront of a grating owing spatial period of tens to hundreds millimeters [28-29], which means the CP with such low spatial frequency is easy to be manufactured.

## 7. Some proposed designs

Here, as shown in Fig. 6, we show a typical four-beam design by using two layer of in-house beam-splitting pulse compressors. Different from tiled grating method where six degrees of freedom need to be precisely controlled, the two gratings here are located up and down in space so that they can share the same optical holder to reduce vibration error between them if independent optical holders are used. Then, we can synchronize and phase-locked the two input beams B_1 and B_2 before this compressor, as shown in the inset of Fig.6, where a delay stage together with the rotation of the compensated plate CP0 is used to synchronize and phase-lock two beams B_1 and B_2. The compensated plate CP0 is also used to compensate the dispersion of the beam splitter BS0.

As for the compressor, G1, G1', G4, G4', G6 and G6' are exactly the same gratings, G2, G2', G3, G3', G5 and G5' are other six identical gratings, R is a whole high reflective mirror with larger size which can be coated with gold or other broadband dielectric film. BS and BS' are 50:50 beam splitters with the same thickness and coating, CP and CP' are compensated plates with same thickness and coating. It need to be noted that BS and BS', CP and CP' can be combined as single beam splitter BS'' or compensated plate CP'' if the manufacture of such large size of transmitted glass plate is possible. In this way, the influence difference induced by BS and BS' or CP and CP' will be reduced during the final four-beam tiled-aperture combining process.

To compensate the spatiotemporal distortion induced by the diffraction wavefront of G2, G3 or G2', G3', the backside of BS or BS' is designed and manufactured. In the same way, the backside of CP or CP' is designed and manufactured to compensate the spatiotemporal distortion induced by diffraction wavefront of G2 and G5, reflective wavefront of R and BS, or diffraction wavefront of G2' and G5', reflective wavefront of R and BS'.

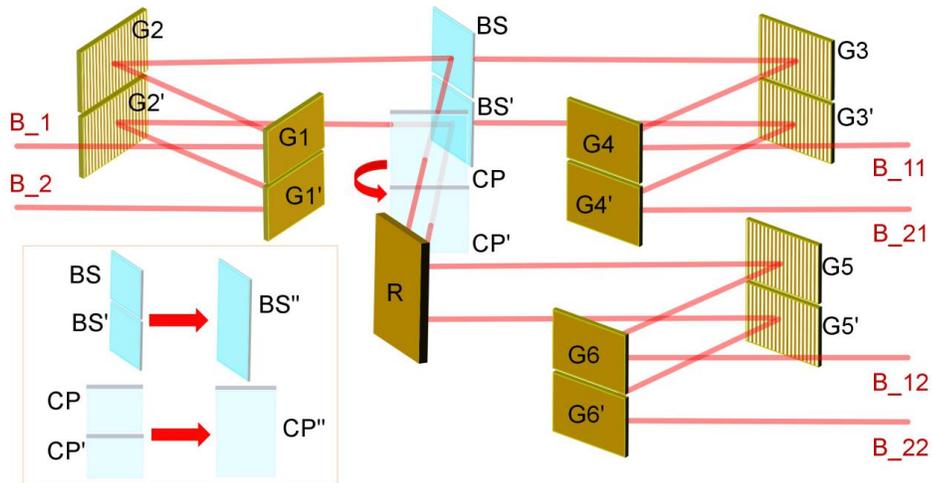

Fig.6 The sketch of proposed two layers in-house beam-splitting pulse compressor together with spatiotemporal coupling compensated plates.

At last, the compressed four laser beams can be combined by using the tiled-aperture combining method [19-26]. In the case, the piston phase, pointing stability, wavefront, and GDD dispersion should be carefully controlled.

## 8. Discussion

Pulse compressor based on diffraction gratings is indispensable for a PW laser facility so far. As for PW-100PW laser system with kJ and even multi-kJ high energy, there is no grating which owes large size, high damage threshold, and broadband diffraction efficiency simultaneously right now. Besides typical four-grating compressor, diamond scheme or special scheme were also developed to limit the number of grating mosaics in ILE, Japan [34-35]. The proposed novel in-house beam-splitting pulse compressor based on typical four-grating compressor design helped to solve this problem well. The novel in-house beam-splitting compressor scheme shows several advantages in comparison to that of the outside beam-splitting pulse compressor shown in Fig.1(b) :

1). It is very clear that the novel in-house design saves expensive gratings and optics with meter size because it shares the first pair of gratings.

2). The two compressors will be more stable, it is because the pointing stability and the piston phase induced by the first pair of gratings will be removed, which is help to achieve and especially maintain the final tiled-aperture combining. It should be noted that every grating owns six degrees of freedom for tiled-grating methods: piston and shift, groove spacing and tilt, rotation and tip [19]. Then, the two gratings are very sensitive due to six degrees of freedom that needs to be stable to maintain the tiled-aperture combing. On the other word, the novel setup will improve the relative stability between the two compressors by 2 times because the first pair of gratings and the second pair of gratings are identical.

3). The sharing of the first pair of gratings also make the two compressed pulses easy to tune and achieve similar spectral phase, which is important for the tiled-aperture combining method [20-21,26]. This is because people do not need to consider the dispersion difference induced by the first pair of gratings. The compensated plate will also exactly balance the material dispersion induced by the added beam splitter.

4). The whole optical setup is more compact which will reduce the whole size of vacuum chamber. Furthermore, it is possible to separate the first pair of gratings into an independent small vacuum chamber, which help to avoid big vacuum chamber which is difficult and expensive to achieve and maintain.

5). The spatial-temporal coupling in the compressor mainly due to the relative large diffraction wavefront distortion of meter-sized gratings, which will decrease the focal intensity and the temporal contrast badly, can be well compensated by using the proposed simple and special compensated plates.

6). As for picosecond PW laser facility which can use dielectric coated gratings for the compressor, the damage threshold ratio between input nanosecond pulse and picosecond or sub-picosecond pulse could be tens to hundreds times. Then, more than two divided beams are possible to be split in the compressor for kJ and even multi-kJ high energy laser pulse output.

## 9. Conclusion

In conclusion, the proposed novel design is proved to be a new pulse compressor that can used to achieve high-energy PW-100PW lasers. This novel design uses the property that the damage threshold of gratings depend on the pulse duration, which means not only femtosecond PW lasers with gold-coated gratings can work well, but also the picosecond or sub-picosecond PW lasers using dielectric coated gratings can work even better with this design. Furthermore, we proposed simple method to compensate the wavefront distortion by adding a compensated glass plate between the second grating and the third grating which will improve the final focal intensity and the temporal contrast. These ultraintense lasers are important tools for ultra-high intense laser-matter interaction researches. Our in-house beam-splitting design also suit well for many important applications, such as photon-photon collisions by PW laser pulses reported recently [36]. These experiments show the requirement of two or more spatial separated while temporal synchronized PW beams. In our design, the independent wide range tunability on pulse duration or chirp of split pulses will also help to extend the experimental conditions. In the future, this novel pulse compressor may also be used together with the upgrade gratings to improve the laser peak power to exawatt level, which allows us to reach much higher focal intensity and explore more frontier researches.

*Acknowledgments*

This work is supported by the National Natural Science Foundation of China (NSFC) (61527821, 61905257, U1930115), the Instrument Developing Project (YZ201538) and the Strategic Priority Research Program (XDB16) of the Chinese Academy of Sciences (CAS), and Shanghai Municipal Science and Technology Major Project (2017SHZDZX02).

The authors would like to thank Prof. Zhaoyang Li from Osaka University for the helpful discussion on spatial-temporal coupling and Prof. Zhaoyang Wei from SIOM for helpful discussion on optics manufacture.